\documentclass[linenumbers]{aastex631}

\usepackage{threeparttable}

\begin{document}
\nolinenumbers
\shorttitle{The outburst of SLX 1746--331 in 2023}
\shortauthors{Peng et al.}
\title{NICER, NuSTAR and Insight-HXMT views to  black hole X-ray binary SLX 1746--331}

\author[0000-0002-5554-1088]{Jing-Qiang Peng\textsuperscript{*}}
\email{pengjq@ihep.ac.cn}
\affiliation{Key Laboratory of Particle Astrophysics, Institute of High Energy Physics, Chinese Academy of Sciences, 100049, Beijing, China}
\affiliation{University of Chinese Academy of Sciences, Chinese Academy of Sciences, 100049, Beijing, China}
\author{Shu Zhang\textsuperscript{*}}
\email{szhang@ihep.ac.cn}
\affiliation{Key Laboratory of Particle Astrophysics, Institute of High Energy Physics, Chinese Academy of Sciences, 100049, Beijing, China}

\author[0000-0001-5160-3344]{Qing-Cang Shui\textsuperscript{*}}
\email{shuiqc@ihep.ac.cn}
\affiliation{Key Laboratory of Particle Astrophysics, Institute of High Energy Physics, Chinese Academy of Sciences, 100049, Beijing, China}
\affiliation{University of Chinese Academy of Sciences, Chinese Academy of Sciences, 100049, Beijing, China}
\author[0000-0001-5586-1017]{Shuang-Nan Zhang}
\affiliation{Key Laboratory of Particle Astrophysics, Institute of High Energy Physics, Chinese Academy of Sciences, 100049, Beijing, China}
\affiliation{University of Chinese Academy of Sciences, Chinese Academy of Sciences, 100049, Beijing, China}

\author[0000-0001-8768-3294]{Yu-Peng Chen}
\affiliation{Key Laboratory of Particle Astrophysics, Institute of High Energy Physics, Chinese Academy of Sciences, 100049, Beijing, China}

\author[0000-0003-3188-9079]{Ling-Da Kong}
\affiliation{Institute f{\"u}r Astronomie und Astrophysik, Kepler Center for Astro and Particle Physics, Eberhard Karls, Universit{\"a}t, Sand 1, D-72076 T{\"u}bingen, Germany}

\author{Zhuo-Li Yu}
\affiliation{Key Laboratory of Particle Astrophysics, Institute of High Energy Physics, Chinese Academy of Sciences, 100049, Beijing, China}
\author[0000-0001-9599-7285]{Long Ji}
\affiliation{School of Physics and Astronomy, Sun Yat-Sen University, Zhuhai, 519082, China}

\author[0000-0002-6454-9540]{Peng-Ju Wang}
\affiliation{Institute f{\"u}r Astronomie und Astrophysik, Kepler Center for Astro and Particle Physics, Eberhard Karls, Universit{\"a}t, Sand 1, D-72076 T{\"u}bingen, Germany}
\author[0000-0002-2749-6638]{Ming-Yu Ge}
\affiliation{Key Laboratory of Particle Astrophysics, Institute of High Energy Physics, Chinese Academy of Sciences, 100049, Beijing, China}
\author[0000-0002-9796-2585]{Jin-Lu Qu}
\affiliation{Key Laboratory of Particle Astrophysics, Institute of High Energy Physics, Chinese Academy of Sciences, 100049, Beijing, China}
\author[0000-0002-2705-4338]{Lian Tao}
\affiliation{Key Laboratory of Particle Astrophysics, Institute of High Energy Physics, Chinese Academy of Sciences, 100049, Beijing, China}

\author[0000-0003-4856-2275]{Zhi Chang}
\affiliation{Key Laboratory of Particle Astrophysics, Institute of High Energy Physics, Chinese Academy of Sciences, 100049, Beijing, China}
\author{Jian Li}
\affiliation{CAS Key Laboratory for Research in Galaxies and Cosmology, Department of Astronomy, University of Science and Technology of China, Hefei 230026, China}
\affiliation{School of Astronomy and Space Science, University of Science and Technology of China, Hefei 230026, China}
\author[0000-0003-2310-8105]{Zhao-sheng Li}
\affiliation{ Key Laboratory of Stars and Interstellar Medium, Xiangtan University, Xiangtan 411105, Hunan, China}

\author{Zhe Yan}
\affiliation{University of Chinese Academy of Sciences, Chinese Academy of Sciences, 100049, Beijing, China}
\affiliation{Yunnan Observatories, Chinese Academy of Sciences, Kunming 650216, China}
\affiliation{Key Laboratory for the Structure and Evolution Celestial Objects, Chinese Academy of Sciences, Kunming 650216, China}
\affiliation{Center for Astronomical Mega-Science, Chinese Academy of Sciences, Beijing 100012, China}



\begin{abstract}
\nolinenumbers

We study the spectral and temporal properties of the black hole X-ray transient binary SLX 1746--331 during the 2023 outburst with NICER, NuSTAR, and Insight-HXMT observations.
Through the joint fitting of the spectra from NICER, NuSTAR, and Insight-HXMT, the spin and inclination angles are measured for the first time as $0.85 \pm 0.03$ and $53\pm 0.5$\textdegree, respectively. Accordingly, the mass of the compact star is updated from the previous lower limit of 3.3$\pm 2.1 M_\odot$ to $5.5\pm 3.6 M_\odot$, which is consistent with $5.2 \pm 4.5M_\odot$ measured with an empirical mass-luminosity correlation of BH samples. With more NICER observations covering the later decay of the outburst, we confirm that the entire outburst was dominated by the disk emissions, and the thermal spectrum follows $F \propto T_{\rm in}^{3.974\pm 0.003}$, 
till a luminosity of over than two magnitudes lower than the maximum of the outburst.

\end{abstract}



\section{Introduction} \label{intro}
A low-mass black hole X-ray binary consists of a companion star with a mass less than 1 $M_\odot$ and a black hole. Due to the lower mass of the companion star, the accreted material is more easily able to fill the Roche lobe, allowing the black hole to accrete material from the companion star through the Roche lobe \citep{1973Shakura}.

For the black hole X-ray transient binary, it remains in a long-term quiescent state with a relatively low accretion rate  \citep{2009Deegan}. As the accreted material in the disk accumulates, it reaches a temperature at which hydrogen can be ionized. Due to thermal and viscous instabilities, the angular momentum transfers outward, the material falls inward, and the accretion rate increases, an X-ray outburst is produced \citep{1995Cannizzo, 2001Lasota,2011Belloni,2016Corral-Santana}.

The outburst of a black hole X-ray binary generally undergoes several different spectral states, including the Low/Hard States (LHS), High/Soft States (HSS), and Intermediate States (IMS)  \citep{2005Belloni,2009Motta}, and these spectral states have distinct positions on the Hardness-Intensity Diagram (HID). The trajectory of the black hole X-ray binary forms a "q" shape on the HID \citep{2001Homan, 2004Fender, 2012Motta}. In the LHS, the emission is primarily dominated by hard X-ray emission from the corona/jet. 
The locations of the accretion disks in the LHS are not clear, and for XTE J1118+480, \cite{2001M} argued that its accretion disk is truncated in the LHS, while \cite{2009R} found a thermal component consistent with a disk at the ISCO, suggesting that its accretion disk is not truncated in the LHS.
NuSTAR results for some canonical sources such as GRS 1915+105 and Cygnus X--1 also show no evidence that the accretion disk is truncated in the LHS \citep{2003M,2015P}.
In the HSS, where thermal emission from the disk dominates and the contribution of non-thermal emission is relatively low (\textless 25\%), the disk has reached its innermost stable circular orbit (ISCO) \citep{1997Esin,2006McClintock,2008Gierlinski}.
The IMS can be further divided into the hard intermediate state (HIMS) and the soft intermediate state (SIMS), where the spectrum of SIMS is softer than that of HIMS \citep{2005Homan}.

Not all black hole X-ray transient binaries exhibit "q" shaped trajectories on the HID, meaning that not all of them undergo a complete outburst from the LMS through the HIMS to the HSS and then back to the LMS through the SIMS. These outbursts are referred to as "failed outbursts," which are mainly classified into two categories: one category includes outbursts that have only evolved to the low hard state or intermediate state but not to the soft state, accounting for approximately 38$\%$ of all outbursts \citep{2004Brocksopp,2009Capitanio,2016Tetarenko}. and the other category is LHS only,  which lacks the beginning of the outburst, which is relatively rare and was in  4U 1630--472, MAXI J0637--430 and SLX 1746--331  \citep{2020Baby,2022Ma,2024p}.

SLX 1746--331  was discovered with the Spacelab 2 XRT in 1985 August and detected by the RASS in 1990 \citep{1988Warwick,1990Skinner}.
\cite{1990Skinner} discovered that it had a very soft spectrum that best fit thermal bremsstrahlung at a temperature of $kT$ = 1.5 keV, leading to the suggestion that SLX 1746-331 might be a black hole candidate.
SLX 1746--331 experienced outbursts in 2003 \citep{2003M,2003R}, 2007 \citep{2007Markwardt}, and 2011 \citep{2011Ozawa}. The observed peak fluxes of these outbursts were approximately 400 mCrab, 100 mCrab, and 40 mCrab, respectively.
\cite{2011D} used data from  Rossi X-ray Timing Explorer (RXTE) and assumed a black hole mass of 10 $M_\odot$, a distance of 5 kpc, and an inclination angle of 60°. By using the $L \propto T_{\rm in}^{4}$ relation, they estimated that $R_{\rm in}$ is approximately 6.7 km, which is about 0.45 times the gravitational radius ($R_{\rm g}$), suggesting this system is either located further away or hosts a black hole with smaller mass.
\cite{2015Yan} estimated the distance of SLX 1746--331 to be about 10.81 $\pm$ 3.52 kpc using data from RXTE.

SLX 1746--331 entered the outburst in 2023 and has been observed by Neutron Star Interior Composition Explorer (NICER) 81 times since March 8, 2023, covering the entire outburst, and has been observed a total of 46 times by Insight-HXMT over 35 days, from March 14th to April 17th, 2023. The previous research mainly based on the early released NICER observations revealed the outburst stayed around ISCO down to a luminosity of as low as 0.03 $L_{\rm Edd}$ \citep{2023P}. 
Nuclear Spectroscopic Telescope Array (NuSTAR) has three observations of SLX 1746--331, carried out during the peak and decay phases of the outburst.
In this letter, we perform a detailed spectral analysis by taking the NuSTAR, complete NICER observations,  and the supplementary  Insight-HXMT observations. 

We report the first measurements on the basic properties of the system,  including the black hole spin, the inclination angle, and the update of the black hole mass.  In Section \ref{obser}, we describe the observations and data reduction. The detailed results are presented in Section \ref{result}. The results are then discussed, and the conclusions are presented in Section \ref{dis}.

\section{Observations and Data reduction}
\label{obser}

\subsection{NICER}

NICER  was launched by the Space X Falcon 9 rocket on 3 June 2017 \citep{2016Gendreau}. NICER has a large effective area and high temporal resolution in soft X-ray band (0.2--12 keV),  which may allow us to better study black body components at low temperatures.

NICER started to observe SLX 1746--331 on March 8, 2023, and covered the entire outburst of the SLX 1746--331.
NICER observations till the MJD 600103 were analyzed and reported in \cite{2023P}. More NICER observations beyond MJD 600126 allow for investigation of the outburst around its later decay phase. 

We use NICERDAS 2022-12-16 V010a software with the latest CALDB xti20221001 for NICER data analysis.
NICER data are reduced using the standard pipeline tool {\tt nicer}l2\footnote{\url{https://heasarc.gsfc.nasa.gov/lheasoft/ftools/headas/nicerl2.html}}. 
We extract light curves using {\tt nicer}l3-lc\footnote{\url{https://heasarc.gsfc.nasa.gov/docs/software/lheasoft/ftools/headas/nicerl3-lc.html}} in 1--6 keV, 6--10 keV and 1--10 keV.
To extract the spectrum, we utilize {\tt nicer}l3-spect\footnote{\url {https://heasarc.gsfc.nasa.gov/docs/software/lheasoft/help/nicerl3-spect.html}}, employing the "{\tt nibackgen3C50}\footnote{\url{https://heasarc.gsfc.nasa.gov/docs/nicer/analysis_threads/background/}}" model to estimate the background for spectral analysis.
For spectrum fitting, we select an energy range of 0.5--10 keV.
Additionally, {\tt nicer}l3-spect automatically applies the systematic error using {\tt niphasyserr}. In the energy range of 0.3-10 kev, the systematic error is about 1.5\%.

\subsection{NuSTAR}

NuSTAR is the first mission to employ focusing telescopes for high-energy X-ray (3 -- 79 keV) imaging of the sky within the electromagnetic spectrum. It was launched on June 13, 2012, at 9 am PDT. \citep{2013Harrison}. 
As shown in Table \ref{observ}, NuSTAR conducted three observations of SLX 1746--331.
For NuSTAR data analysis, we employ NuSTARDAS v2.1.2 software with the calibration database (CALDB 20230613).
We extract NuSTAR-filtered data using the standard pipeline program {\tt nupipeline},  
The spectrum was extracted from a 120$''$ circle region centered on the source and the background was generated from a 60$''$ circle region away from the source. We utilized FPMA and FPMB 3--50 keV data for spectral analysis.

\subsection{Insight-HXMT}
Insight-HXMT is the first Chinese X-ray astronomy satellite, which was successfully launched on 2017 June 15 \citep{2014Zhang, 2018Zhang, 2020Zhang}.  It carries three scientific payloads: the low energy X-ray telescope (LE, SCD detector, 1--15 keV, 384 $\rm cm^{2}$, \citealt{2020Chen}), the medium energy X-ray telescope (ME, Si-PIN detector, 5--35 keV, 952 $\rm cm^{2}$, \citealt{2020Cao} ), and the high energy X-ray telescope (HE, phoswich NaI(CsI), 20--250 keV, 5100 $\rm cm^{2}$, \citealt{2020Liu}).

Insight-HXMT started to observe SLX 1746--331 at the peak of its outburst on March 14, 2023 and continued until May 19, 2023. As shown in Table \ref{observ}, only the Insight-HXMT observations simultaneous with NICER and NuSTAR are adopted for joint spectral analysis.
We extract the data from LE and ME using the Insight-HXMT Data Analysis software {\tt{HXMTDAS v2.05}}. 
The data are filtered with the criteria recommended by the Insight-HXMT Data Reduction Guide {\tt v2.05} \footnote{\url{http://hxmtweb.ihep.ac.cn/SoftDoc/648.jhtml}}.
Due to the count rate of high-energy photons in SLX 1746--331, the energy bands considered for spectral analysis are LE 2--8 keV and ME 8--25 keV. One percent systematic error is added to data \citep{2020Liao}, and errors are estimated via  Markov Chain Monte-Carlo (MCMC) chains with a length of 20000.

\section{Results}
\label{result}

\subsection{Light curve and Hardness-intensity diagram}
\label{light curve}

\begin{figure}
	\centering
	\includegraphics[angle=0,scale=0.6]{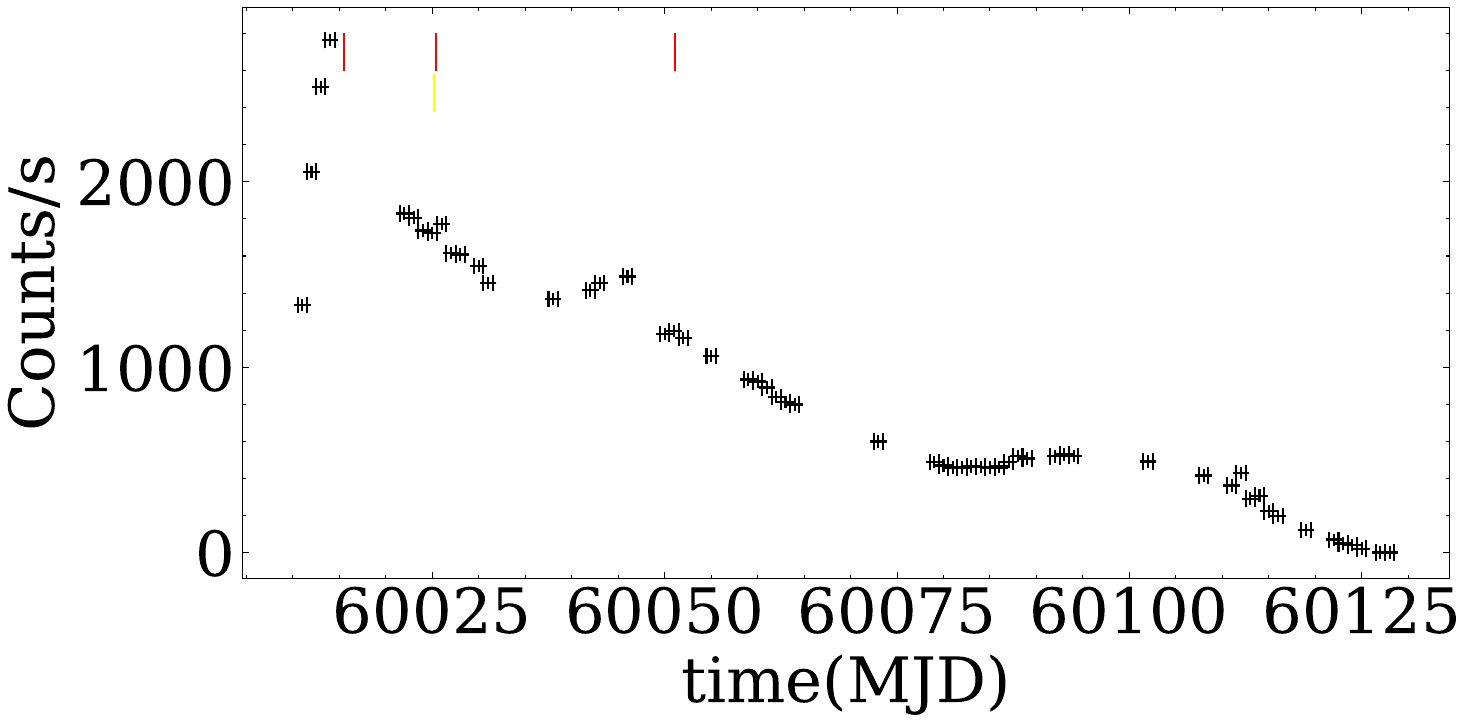}
	\caption{The light curve of SLX 1746--331 for NICER 1--10 keV during 2023 outburst. The three red bars represent the three NuSTAR observations. The yellow bar represents the Insight-HXMT observation. }
	\label{lcurve}
\end{figure}

\begin{figure}
	\centering
	\includegraphics[angle=0,scale=0.31]{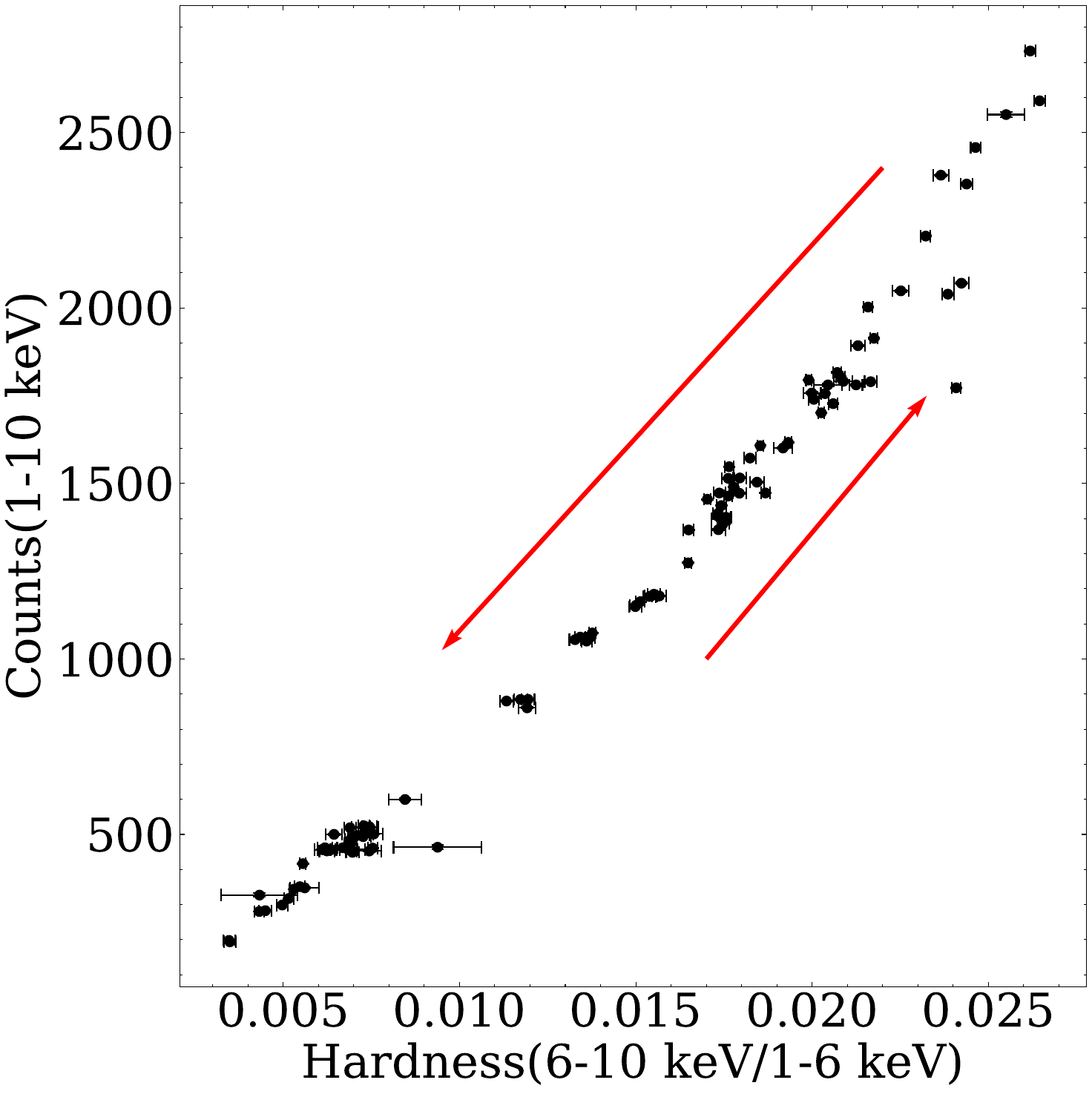}
	\caption{The NICER hardness-intensity diagram of SLX 1746--331, where the hardness is defined as the ratio of 6--10 keV to 1--6 keV count rate. }
	\label{HID}
\end{figure}

\begin{table}[htbp]
    \centering
		\caption{NICER, NuSTAR and Insight-HXMT    observations of SLX 1746--331  during the 2023 outburst. }
		\begin{tabular}{cccccc}
		%\begin{tabular}{|c|m{2cm}|}
		  \hline
		   \hline
		   NICER & Observed date & Exposure Time \\ObsID& (MJD) & (s) \\ \hline
      6203700105 &60015.05 & 766 \\ 
      6203700110 &60025.30 & 3888 \\
      6203700122 &60051.66 & 1797 \\
       \hline \hline
         NuSTAR &  Observed date & FPMA exposure time & FPMB exposure time
         \\ ObsID& (MJD) & (s) & (s) 
         \\ \hline
       80802346002 &  60015.45 &17180& 17790 \\ 
       80802346004 &  60025.36 &23440& 23690 \\ 
       80802346006 &  60051.08 &24290& 24600 \\ 
       		  \hline \hline
           
         Insight-HXMT &  Observed date & LE exposure time & ME exposure time
         \\ ObsID& (MJD) & (s) & (s) 
       \\ \hline
       P051436300803 & 60025.22& 284.3 & 964.2  \\ 
    \hline
        \label{observ} &     

    \end{tabular}

\end{table}

SLX 1746--331 entered the outburst on March 8, 2023. As shown in Figure \ref{lcurve}, NICER began observing it on March 8, 2023 (MJD 60011), covering the entire outburst. The outburst manifests with a profile of "fast rise and slow decay", which is a typical characteristic of those outbursts observed in most black hole X-ray binary systems.  The three short red lines in Figure \ref{lcurve} represent NuSTAR observations, and the yellow line denotes Insight-HXMT observation (Table \ref{observ}).

We extract the 1–6 keV, 6–10 keV, and 1–10 keV light curves of NICER to construct the Hardness-Intensity Diagram (HID) of SLX 1746-–331. The hardness is defined as the count rate ratio of 6–10 keV to 1–6 keV, while the intensity is the count rate ratio of 1–10 keV.  As shown in Figure \ref{HID},  the red arrows in the HID represent the trajectory of SLX 1746-–331, showing that the outburst of SLX 1746--331 commences and evolves to a soft state, with no trend for a transition to the hard state during the entire outburst.

\subsection{The spectral analysis}
\label{parameters}

\begin{figure}
	\centering
	\includegraphics[angle=0,scale=0.4]{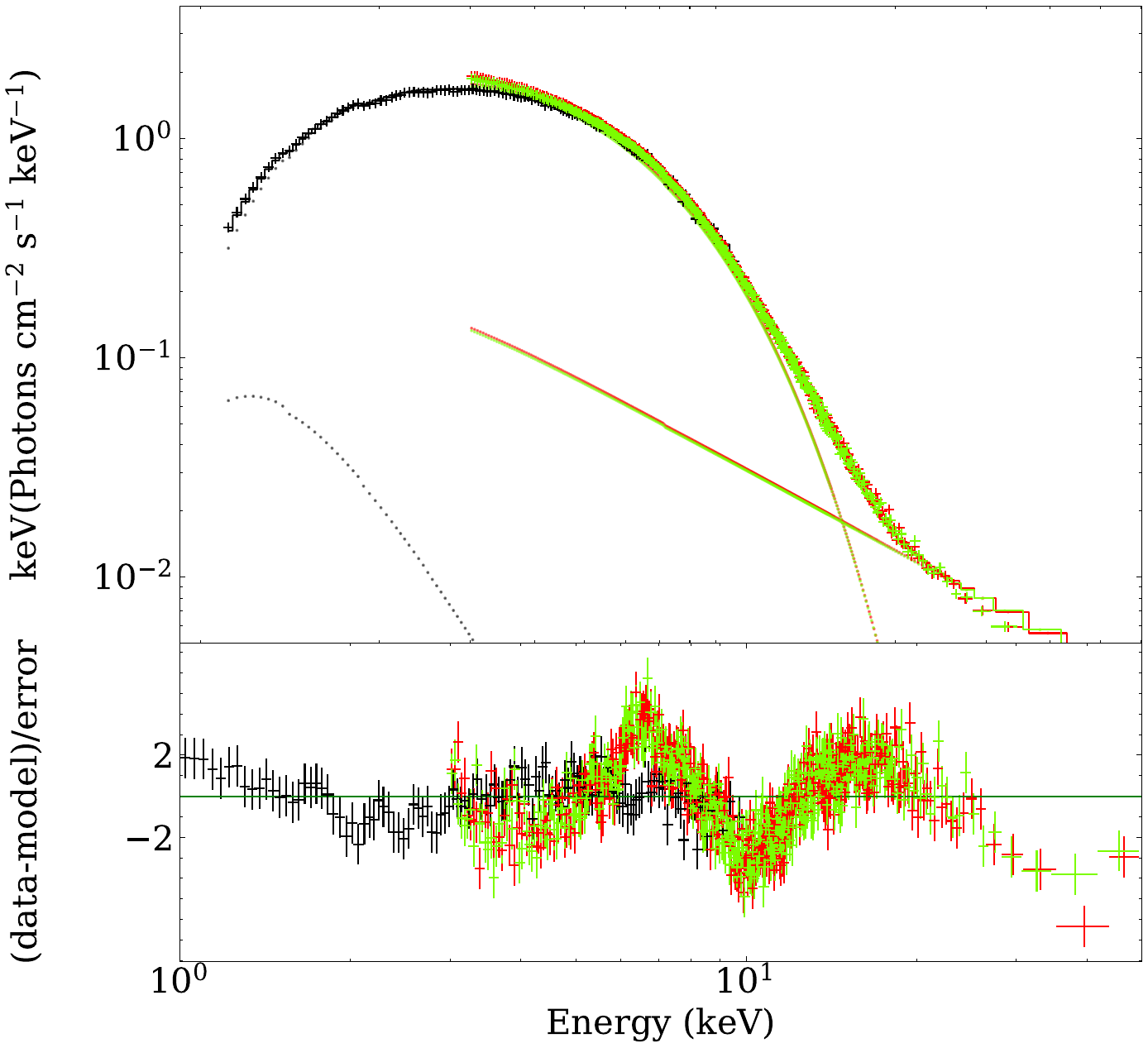}
	\caption{The simultaneous broadband spectrum of SLX 1746–331 is observed from NICER (black), NuSTAR/FPMA (red), and NuSTAR/FPMB (green),}
	\label{re}
\end{figure}

\begin{figure}
	\centering
	\includegraphics[angle=0,scale=0.4]{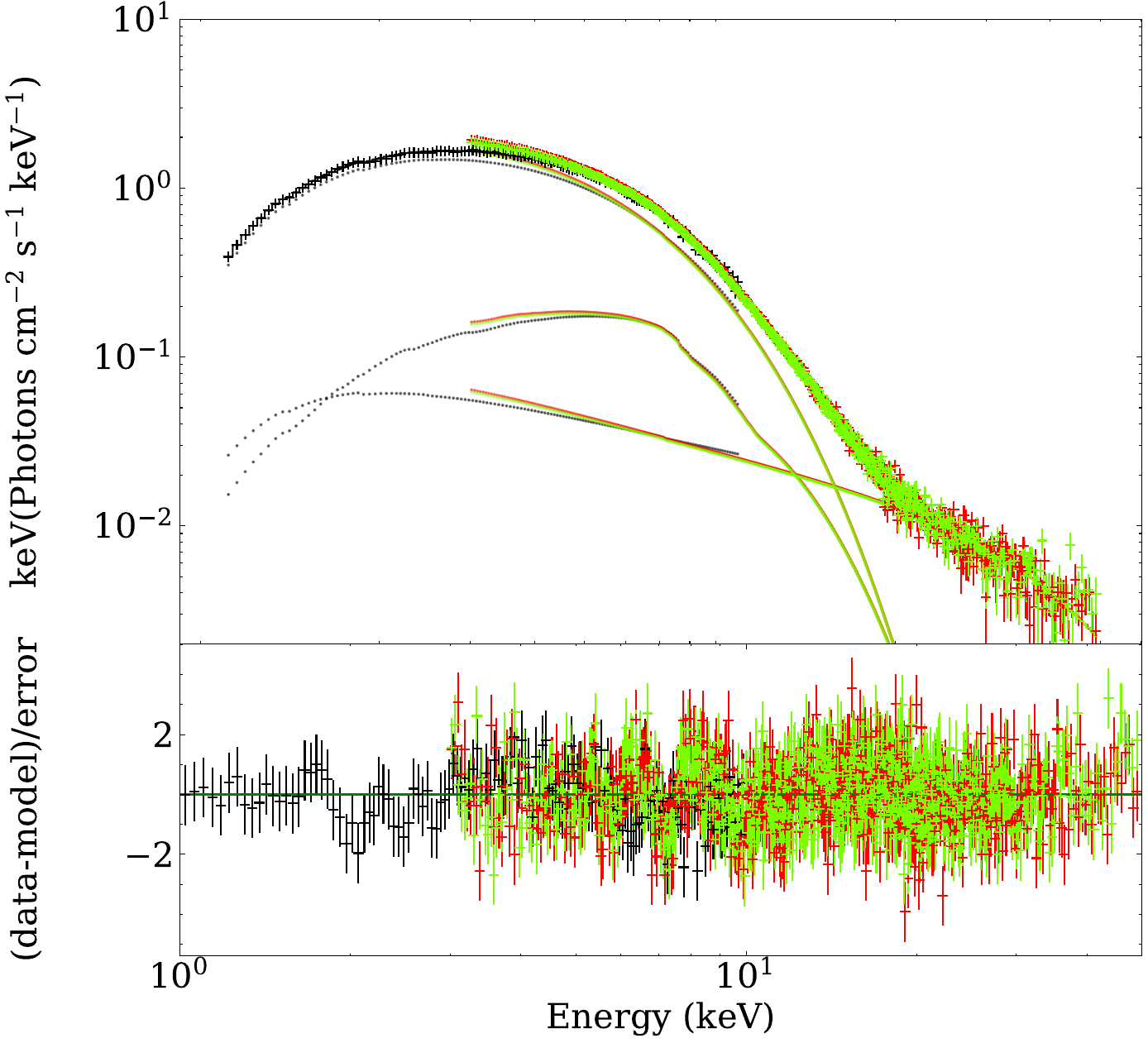}
	\caption{The simultaneous broadband spectrum of SLX 1746–331 is observed from NICER (black), NuSTAR/FPMA (red), and NuSTAR/FPMB (green),}
	\label{fit}
\end{figure}

\begin{table}[htbp]
    \centering
    \renewcommand{\arraystretch}{1.5}
     \setlength{\tabcolsep}{5pt}
		\caption{The results of spectral fitting the NICER+NuSTAR+Insight-HXMT data for Model M2.}
        \huge
		\resizebox{12cm}{!}{
		\begin{threeparttable}
	\begin{tabular}{ccccccc}
		%\begin{tabular}{|c|m{2cm}|}
		
		  \hline
		   \hline
        Model & Parameter &80802346002&80802346004&80802346006
       \\ \hline
tbabs& $N_{\rm H}[10^{22} \rm cm^{-2}]$&$0.76^{+0.01}_{-0.01}$&$0.86^{+0.01}_{-0.01}$&$0.88^{+0.01}_{-0.05}$\\
\hline
ezdiskbb&$T_{\rm max}$&$1.69_{-0.02}^{+0.01}$&$1.52^{+0.01}_{-0.01}$&$1.34^{+0.02}_{-0.01}$ \\
&norm&$17.8_{-0.2}^{+0.1}$&$16.8_{-0.2}^{+0.3}$&$17.4^{+0.5}_{-1.1}$\\
&flux[$10^{-8}$~erg~s$^{-1}$~cm$^{-2}$]&$1.36_{-0.02}^{+0.03}$&$0.73_{-0.02}^{+0.01}$&$0.52_{-0.02}^{+0.02}$\\
\hline
relxillNS&$a$&$0.82_{-0.06}^{+0.07}$&$0.88_{-0.05}^{+0.04}$&$0.85_{-0.01}^{+0.03}$ \\
&$i$ [\textdegree]&$51.2_{-1.3}^{+1.1}$&$52.6_{-0.6}^{+0.5}$&$53.1_{-0.7}^{+0.3}$\\
&logxi&$3.32_{-0.03}^{+0.09}$&$3.05_{-0.03}^{+0.05}$&$3.48_{-0.03}^{+0.01}$\\
&$A_{fe}$&$0.87_{-0.04}^{+0.51}$&$0.86_{-0.14}^{+0.15}$&$1.3^{*}$\\
&norm[$10^{-3}$]&$3.1_{-0.2}^{+0.3}$&$1.6_{-0.2}^{+0.1}$&$1.2_{-0.1}^{+0.0}$\\
\hline

nthcomp&$\Gamma$&$2.31_{-0.03}^{+0.06}$&$3.27_{-0.03}^{+0.02}$&$3.47_{-0.07}^{+0.11}$\\
&$kT_{\rm e}$[keV]&$17.40_{-2.33}^{+3.79}$&$6.13_{-0.88}^{+1.15}$&$13.87_{-4.03}^{+10.29}$\\
&norm&$0.41_{-0.04}^{+0.08}$&$0.81_{-0.11}^{+0.14}$&$1.12_{-0.16}^{+0.41}$\\
\hline
constant
&con[LE]&&$1.25_{-0.05}^{+0.04}$&\\
&con[ME]&&$0.79_{-0.11}^{+0.12}$&\\
&con[NFPMA]&$1.45_{-0.01}^{+0.01}$&$1.38_{-0.04}^{+0.03}$&$1.31_{-0.02}^{+0.04}$\\
&con[NFPMB]&$1.37_{-0.02}^{+0.01}$&$1.31_{-0.03}^{+0.03}$&$1.19_{-0.04}^{+0.02}$\\
\hline
&$\chi^2$/(d.o.f.)&1.09& 1.07&1.24\\
\hline
	\label{parameter} &
    \end{tabular}
\begin{tablenotes}[para,flushleft] 
        \item Note:\\
        **We fixed the iron abundance by the results of the joint fit to 1.3
     \end{tablenotes} 
     
\end{threeparttable}} 

\end{table}

\begin{figure}
	\centering
	\includegraphics[angle=0,scale=0.3]{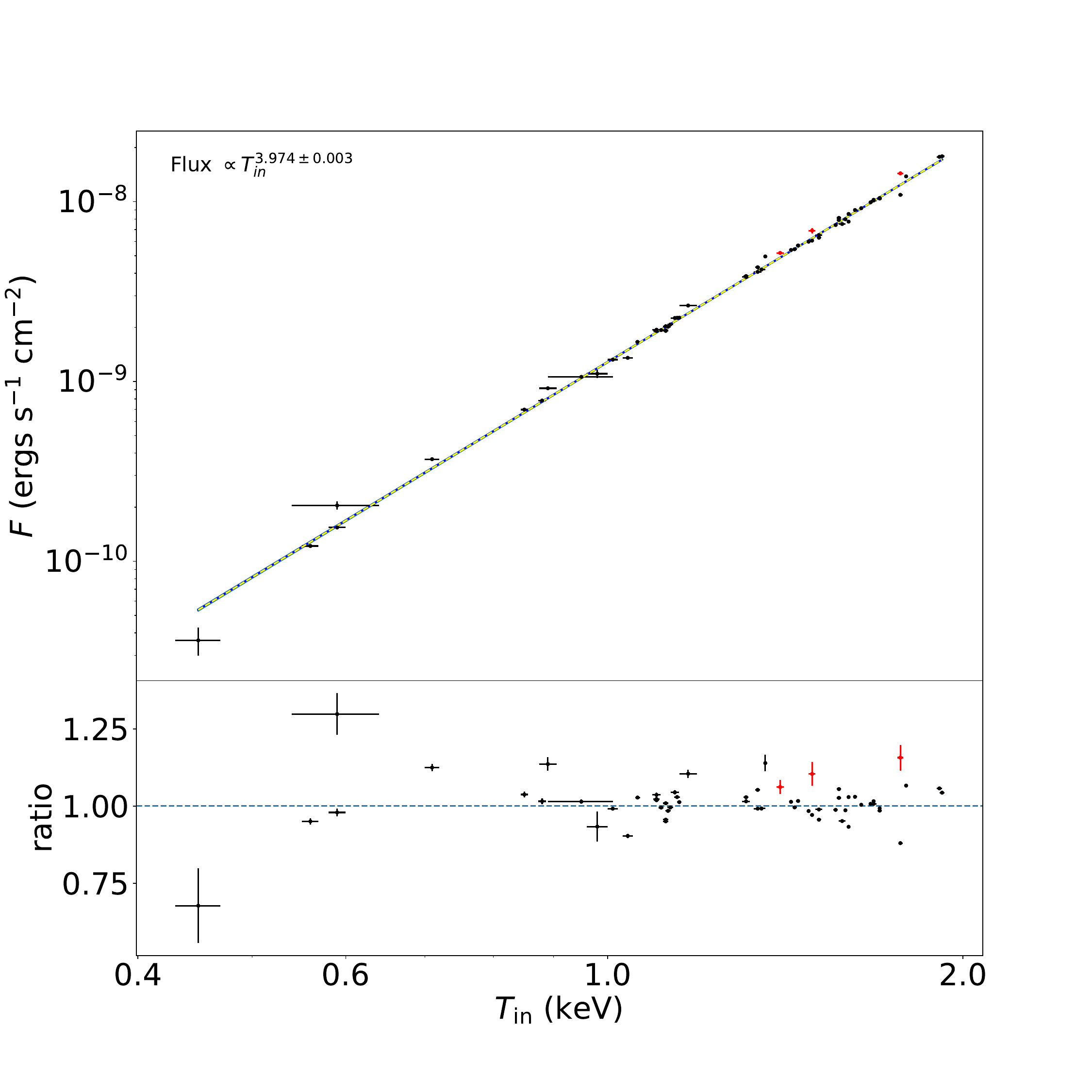}
	\caption{The disk unabsorbed flux (1--100 keV) versus the disk inner temperature ($T_{\rm in}$). Flux is found to correlate with temperature in the form of proportional $T^{3.974\pm 0.003
 }_{\rm in}$. The red points represent the results of the joint fit.}
	\label{T4}
\end{figure}

For the NICER data, we fit the 0.5--10 keV spectra using a model that includes {\tt tbabs} to account for interstellar absorption \citep{2000Wilms}, considering photoelectric cross-sections provided by \cite{1996Verner}, and {\tt diskbb} for the multi-temperature blackbodspy of the accretion disk \citep{1984Mitsuda}.  So our model M1 is {\tt tbabs*diskbb} and the flux of the disk in the 1--100 keV is estimated with {\tt cflux}.

 As shown in Figure \ref{lcurve} and Table \ref{observ}, there are three NuSTAR observations of SLX 1746--331, all of which have nearly simultaneous NICER observations. Additionally, one of these observations also had nearly simultaneous observations from Insight-HXMT.
Since the spectrum of SLX 1746--331 is soft and the high-energy photon counts are relatively low, we conducted spectral analysis of SLX 1746--331 using nearly simultaneous data from NICER in  0.5--10 keV, NuSTAR FPMA and FPMB in 3--50 keV, and Insight-HXMT LE in 2--8 keV and ME in 8--25 keV.
The first trial for the spectral model is {\tt constant*tbabs*(ezdiskbb+powerlaw)}, where {\tt ezdiskbb} is a model of multiple blackbody disks with zero-torque inner boundary \citep{2005Zi}. In order to take into account the differences between the different spectral indices of NICER and NuSTAR, we allowed the temperature of the disk and spectral index to vary.
We find a poor-fitting result with $\chi^2$/(d.o.f)=3358.67/1345=2.5, and there is a significant reflection component in the residuals(Figure \ref{re}). So we replaced {\tt powerlaw} with {\tt relxill} to fit the reflection component in the spectrum \cite{2016Dauser}, which results in an improving but not yet perfect fitting with $\chi^2$/(d.o.f)=1896.32/1340=1.4. Then we add the {\tt nthcomp} model to improve the fit, where the  $\Gamma$ in {\tt relxill} is linked to $\Gamma$ in {\tt nthcomp}. The reflection fraction is fixed to -1, so we only consider the reflection emission [$\chi^2$/(d.o.f)=1746.27/1338=1.3]. 
Since the fit is not improved much, we replace  { \tt relxill} with {\tt relxillNS}, where the reflection fraction is fixed to -1 and  $kT_{\rm bb}$ in {\tt relxillNS} linked to $T_{\rm in}$ in {\tt nthcomp}.  $R_{in}$ is fixed to -1, i.e., the disk is in the ISCO, emission index and truncation radius of 3 and 15, respectively. Finally the fit is largely improved [$\chi^2$/(d.o.f)=1462.19/1339=1.09]. Therefore our fitting model M2 is:{\tt constant*tbabs*(ezdiskbb+nthcomp+relxillNS)}. The flux of the disk in the 1--100 keV is estimated with {\tt cflux}.  Figure \ref{fit} shows the spectral fitting of SLX 1746--331, and the spectral parameters are shown in Table \ref{parameter}, where the black hole spin and the inclination angle of the system are measured for the first time with the averaged values of  $0.85 \pm 0.03$ and $53\pm 0.5$\textdegree, respectively. 

We also try to fit the energy spectrum by replacing {\tt ezdiskbb} with {\tt kerrbb}, because {\tt kerrbb} has spin, inclination, and other parameters. We link the spin and inclination with those of the {\tt relxillNS} model and fix the distance to 10, the mass to 5 and, according to the requirements of the model, fix the norm to 1. We find that the spin and inclination obtained in the fit are the same as those obtained with the model we used previously. We find that the spin and inclination are basically the same as what we obtained with model 2.
To obtain more robust results, we jointly fit the spectra from the three separate days,  with parameters of black hole spin, inner disk inclination, and iron abundance linked among these observations, and obtain spin, inclination and iron abundance as $0.86_{-0.02}^{+0.03}$, $54.4_{-1.1}^{+0.8}$ and $1.31_{-0.04}^{+0.07}$ respectively. These parameter estimations are consistent with previous results.

The disk flux versus the inner disk temperature ($T_{\rm in}$) is plotted in Figure \ref{T4}, and a fit with power law results in $\alpha$ around  3.974$\pm$0.003. 
This finding suggests that the disk is around the ISCO during the whole outburst. The highest and lowest fluxes of the disk differ by a factor of 500.

\subsection{Estimating the mass of compact objects}
\label{estimation}
As shown in Figure \ref{T4}, the disk flux is always proportional to the fourth power of the temperature of the inner disk during the 2023 outburst of SLX 1746--331.
For the standard thin disk,
$L_{\rm disk}\approx 4\pi R_{\rm in}^{2}\sigma T_{\rm in}^{4}$, where $R_{\rm in}$, $T_{\rm in}$, and $\sigma$ representing the inner radius, the temperature of the disk and Stefan-Boltzmann constant respectively.
This indicates that the disk is around the ISCO. Note that the standard disk applies to a luminosity range of 0.001 $L_{\rm Edd}$ $\lesssim $ $L_{\rm disk}$ $\lesssim$0.3 $L_{\rm Edd}$ \citep{2010Steiner,2013Salvesen,2015Garcia,2023Draghis}, where $L_{\rm Edd}$ is the Eddington luminosity ($L_{\rm Edd}=1.26\times 10^{38} \left(\frac{M}{M_\odot}\right)\text{erg/s}$).
As shown in Figure \ref{T4}, we find that the maximum flux for the disk is about 491 times greater than the lowest flux. This luminosity coverage exceeds that commonly associated with standard disks (0.001 $L_{\rm Edd}$ $\lesssim $ $L_{\rm disk}$ $\lesssim$0.3 $L_{\rm Edd}$).
So we speculate that the peak flux most likely has reached a value around 0.3 $L_{\rm Edd}$, and hence we use the flux and luminosity conversion in X-ray binaries to estimate the mass of the compact object. For a disk blackbody emission, the luminosity of the disk can be approximated as 
$L_{\rm disk} \approx \frac{2\pi F d^2}{\rm cos\theta}$, where the distance $d = 10.81 \pm$3.52 kpc \citep{2015Yan}, and thus a peak luminosity is $L_{\rm disk}=\frac{1.24 \pm 0.81\times 10^{38}} {\rm cos\theta}$ erg/s.  As shown in Table \ref{parameter}, the joint spectral fittings with NuSTAR, NICER, and Insight-HXMT,  result in an average inclination angle $53\pm 0.5$\textdegree.  Accordingly, the mass of the compact object could be estimated as  $5.5\pm 3.6 M_\odot$.

In section \ref{parameters}, we used the M1 model to fit the spectrum of NICER and obtained a normalization value of approximately $70 \pm 10$ for the disk.  For the normalization of the disk, $\text{norm} = \left(\frac{\text{Rin}}{\text{D}_{10}}\right)^2 \cos\theta$, where $R_{\rm in}$ is the inner disk radius, $D_{\rm 10}$ is the distance of the source in units of 10 kpc (1.08 $\pm$ 0.352 kpc), and $\theta$ is the inclination angle ($53\pm 0.5$\textdegree). we estimate the inner disk radius $R_{\rm in}$ to be 11.7 $\pm$ 3.9 km. 
For M2 we estimate the inner disk radius $R_{\rm in}$ to be 5.8 $\pm$ 1.9 km. For a Kerr black hole, with a measured spin a=$0.85\pm 0.03$  and a mass of the compact object about $5.5\pm 3.6 M_\odot$, $R_{\rm Isco}$ gives $21.3_{-14.5}^{+16.2}$ km. Within the error range, the inner disk radius is consistently estimated in two different manners.

\section{discussion and conclusion}
\label{dis}

We have joint spectral analyses of the 2023 SLX 1746-331 outburst using data from NICER, NuSTAR, and Insight-HXMT, which have been carried out, and for the first time, the spin and inclination of SLX 1746-331 are measured approximately as $0.85\pm 0.03$ and $53\pm 0.5$\textdegree, respectively. By taking a peak outburst flux of 0.3 $L_{\rm Edd}$ and the measured inclination, the mass of the compact object is derived as $5.5\pm 3.6 M_\odot$. Additionally, with the complete NICER observations, we also updated that during the entire outburst, is found to have its inner disk staying around  ISCO even at the later decay phase.

The previous outburst of SLX 1746--331 exhibited characteristics that was identified as a black hole transient X-ray binary \citep{1990Skinner,1996White,2003Homan}.
\cite{2015Yan} reported a distance of about 10.81$\pm 3.52$ kpc using RXTE data, and \cite{2024p} estimated a lower limit of $3.3\pm 2.1 M_\odot$ for the mass of the compact object by assuming 0.3 $L_{\rm Edd}$ of the outburst peak flux, and  $5.2 \pm 4.5M_\odot$ from an empirical correlation between disk temperature and black hole mass. 
The large mass uncertainty is due to the shortage of precise measurements of the inclination of the system and the distance. By performing the joint spectral fittings with NICER, NuSTAR, and Insight-HXMT, the disk inclination is firstly precisely measured, which results in an updated mass measurement of the  $5.5\pm 3.6 M_\odot$, well consistent with that derived from the previous empirical mass-temperature correlation \citep{2023P}. Obviously, mass can be further constrained in the future if the source distance can be more precisely measured.

For the 2023 outburst of SLX 1746--331, the analytical results of the spectrum of NICER, and the HID indicates that it remained mostly in a soft state. However, we have observed a significant reflection component in the three NuSTAR observations, where the disk is most likely illuminated by itself at the inner part. In the mean while, the non-thermal emissions contribute about 20\% to the total. 
It is generally believed that there is a hard component in the high soft state, but its proportion is relatively low (\textless25\%).  \cite{2009L} argued that the evaporation rate that corresponds to the corona production is related to the distance of the inner disk for the central black hole.  As shown in their evaporation curve, the distant disk can be truncated by evaporation due to insufficient mass supply. When the accretion rate increases to a maximum value (0.027 times the Eddington accretion rate), evaporation cannot evacuate any disk region because the mass flow in the disk is greater than the evaporation rate. As a result, the disk extends into the ISCO, the corona covering the disk becomes very weak.  For SLX 1746--331, no low hard state and intermediate states may mean the evaporation is very low during the entire outburst.

The 2023 outburst of SLX 1746--331 differs from those of normal black hole transient X-ray binaries in their evolution patterns in HID: instead of a low-hard state toward a soft state, SLX 1746--331 stayed around the ISCO during the whole outburst. Although such an outburst phenomena is relatively rare, we notice that so far a similar case was observed in the outburst of another black hole X-ray binary system MAXI J0637--430 \citep{2022Ma}. 
\cite{2023P} compared SLX 1746--331 with MAXI J0637--430, and suggested that such peculiar outburst could be related to the special configuration of the binary system. For example, a small compact object mass together with a rather tight orbital period could lead to a relatively smaller disk size which could own a weaker large-scale magnetic field \citep{2021Cao}. As discussed in \cite{2023P} and \cite{2021Cao}, a weak magnetic field born out of a smaller disk can in principle lead the state transition toward HSS occurred at a lower luminosity level.  Unlike MAXI J0637--430, which had a peak luminosity of about 0.08 $L_{\rm Edd}$ and a soft-to-hard transition at about 0.007 $L_{\rm Edd}$, SLX 1746--331 remains in the soft state,  with no state transition observed even at a luminosity \textless 500 times lower than the peak luminosity of the outburst. 

If takes the maximum luminosity of 0.3$L_{\rm Edd}$ for the outburst, the luminosity with which the disk remains around ISCO could be as low as 0.0006 $L_{\rm Edd}$, lower than 0.001$L_{\rm Edd}$ established from previous numerical simulations and observations \citep{2008R,2010Steiner,2013Salvesen}.
To further understand such a peculiar outburst, in future there two key issues in probing this binary system apart from the precise measurement of the source distance. One is the orbital period and the other is the state transitions. Since MAXI J0637--430 was already detected with so far the tightest orbital period of $2.2^{+0.8}_{-0.6}$ hrs, given the non-detection of a return of the spectral state to the hard state for SLX 1746--331 during the outburst decay phase, we expect an orbit even tighter than that of MAXI J0637--430. To catch a state transition either toward HSS in othe utburst rising phase or away from HSS in the decay phase, would also be crucial to our pining down the puzzle.  Actually such a task has already served as one of the key sciences of the Einstein probe, which was successfully launched on January 9, 2024 to discover and observe sources exhibiting activity at very low luminosity. \citep{2018Yuan}.

%% Also note that the akcnowlodgment environment does not support long amounts of text. If you have a lot of people and institutions to acknowledge, do not use this command. Instead, create a new \section{Acknowledgments}.
\begin{acknowledgments}
This work is supported by the National Key R\&D Program of China (2021YFA0718500), the National Natural Science Foundation of China under grants No. 12333007, U1838202,  U2038101, U1938103, 12273030, U1938107 and 12027803.
This work made use of data and software from the Insight-HXMT mission, a project funded by China National Space Administration (CNSA) and the Chinese Academy of Sciences(CAS). This work was partially supported by International Partnership Program of Chinese Academy of Sciences (Grant No.113111KYSB20190020).
This research has made use of software provided by of data obtained from the High Energy Astrophysics Science Archive Research Center (HEASARC), provided by NASA’s Goddard Space Flight Center.
L. D. Kong is grateful for the financial support provided by the Sino-German (CSC-DAAD) Postdoc Scholarship Program (91839752).

\end{acknowledgments}

\bibliography{sample631}{}
\bibliographystyle{aasjournal}



\end{document}